\documentstyle[12pt,psfig]{article}
\newcommand{\be}{\begin{equation}}
\newcommand{\ee}{\end{equation}}

\newcommand{\bea}{\begin{eqnarray}}
\newcommand{\eea}{\end{eqnarray}}

\begin{document}
\title{Probes for the early reaction dynamics of heavy-ion collisions
at AGS and SPS}
\author{
{\bf A. Dumitru, J. Brachmann, M. Bleicher,}\\
{\bf J.A. Maruhn, H. St\"ocker, W. Greiner}
\\[0.3cm]
{\small Institut f\"ur Theoretische Physik der J.W.Goethe Universit\"at}\\
{\small Postfach 111932, D-60054 Frankfurt a.M., Germany}
\\[1cm]
{\small Proceedings of the}\\
{\small Workshop on Hydrodynamics at ECT*}\\
{\small Trento (Italy), Mai 12 - 17, 1997}
\\[2cm]
}
%\date{\today}
\maketitle   
\begin{abstract}
We discuss the early evolution of ultrarelativistic heavy-ion collisions
within a multi-fluid dynamical model. In particular, we show that
due to the finite mean-free path of the particles compression shock waves are
smeared out considerably as compared to the one-fluid limit. Also, the
maximal energy density of the baryons is much lower. We discuss the time
scale of kinetic equilibration of the baryons in the central region and
its relevance for directed flow. Finally, thermal emission of direct photons
from the fluid of produced particles is calculated within the three-fluid
model and two other simple expansion models. It is shown that the
transverse momentum and rapidity spectra of photons give clue to the
cooling law and the early rapidity distribution of the photon source.
\end{abstract}
\vspace*{-21cm}
\begin{flushright}
%nucl-th/9703044
\end{flushright}
\newpage
 
\section{Introduction}
 
We discuss a three fluid hydrodynamical model for ultrarelativistic
heavy-ion collisions in the energy range from BNL-AGS to CERN-SPS.
The three fluids are introduced to separate the various rapidity regions
observed in high-energy $pp$-collisions~\cite{refppdNdy}. There it was
found that the baryon charge essentially remains close to projectile
resp.\ target rapidity and the energy loss due to particle production is
transfered to midrapidity. Assuming that the initial (!) stage of
ultrarelativistic heavy-ion reactions is essentially an incoherent
superposition of binary $NN$-collisions leads to a different picture of
the early reaction dynamics as compared to one-fluid hydrodynamics.
The projectile and target nucleons (as well as the newly produced particles)
do not thermalize instantaneously (in the first interaction) but
have to undergo several scatterings until kinetic equilibrium is eventually
established.

If the above-mentioned picture of the compressional stage of the
heavy-ion reaction is correct, it is natural to introduce three fluids
corresponding to the nucleons of the projectile and target (fluids one and
two), and to the particles produced around midrapidity (fluid 3). The fluids
have to be coupled via local friction forces leading to energy- and
momentum exchange. The third fluid is, of course, absent initially and is
produced in the course of the collision due to binary collisions between
the nucleons of projectile and target. Presently, the rescattering
of the produced particles with the nucleons is not taken into account.
The interactions between fluids one and two are derived from
data on $NN$-collisions. We employ the parametrization of ref.\
\cite{2fluid}. A detailed discussion of our model as well as results
on baryon stopping, kinetic equilibration, directed and radial baryon flow
at AGS were presented in refs.\ \cite{3fluid}. In contrast to previous  
multi-fluid models \cite{previous}, in our model
\begin{enumerate}
\item three fluids, corresponding to projectile, target, and produced   
particles
\item local interactions between the fluids which are based on
$NN$-interactions
\item $(3+1)$-dimensional propagation of all three fluids in space-time
\end{enumerate}
are implemented.
 
\section{Compression shocks}
\begin{figure}[htb]
%\vspace*{-1.0cm}   
\centerline{\hbox{ 
\psfig{figure=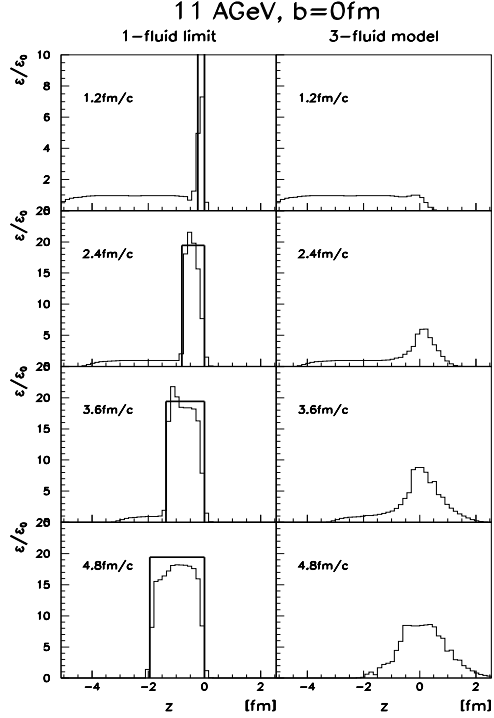,height=10cm,width=7cm}}}
\vspace*{-.5cm}
\caption[]{
Energy density profile of the projectile fluid along the beam axis (in the
center of the reaction, i.e.\ $x=y=0$)
in one-fluid (left) and three-fluid (right) hydrodynamics as a function of
CMS-time.}
\label{fig1}
\end{figure}
In (ideal) one-fluid hydrodynamics the projectile and target nucleons stop
completely (in the center-of-momentum frame) when they touch. Therefore, two
shock discontinuities are created \cite{Scheid} that propagate outwards if the
velocity of the nucleons is supersonic and if matter is thermodynamically
normal, cf.\ ref.\ \cite{RiShock} and references therein. Matter between the
shock
waves is completely at rest (in the CMS) and reaches the maximal possible
energy and baryon density that is compatible with local energy-momentum and
baryon number conservation \cite{Taub,RiShock}.
In fig.\ \ref{fig1} we show that the energy density of the shocked
projectile matter
and the velocity of the shock wave in the three-dimensional calculation 
agree reasonably well with the values obtained in the
semi-analytical one-dimensional
shock model (indicated by the box) \cite{Taub,RiShock} with our
equation-of-state, which
treats the baryonic fluids as a non-relativistic ideal gas
with compression energy \cite{Scheid}:
\be\label{EoS}
p(\epsilon,n) = \frac{2}{3} (\epsilon-E_c n) +p_c \quad.
\ee
For the compression energy, we employ the ansatz
\be\label{comprene}
E_c=\frac{k_c}{18n n_0} (n-n_0)^2 +m_N +W_0 \quad,
n_0 \approx0.16~fm^{-3}\quad,
\ee
so that the compressional pressure $p_c$ is
\be
p_c = -\frac{dE_c}{dn^{-1}}=n^2\frac{dE_c}{dn}=\frac{k_c}{18n_0}
(n^2-n_0^2) \quad.
\ee
However, one also observes that the full-step SHASTA \cite{SHASTA}
overshoots the correct value for the energy density of the compressed
matter at the shockfront. This could probably be cured using a half-step 
treatment \cite{RiShock}, which however requires considerably higher
computing times and thus can not be easily implemented
in multi-dimensional, multi-fluid models.
Once the shock wave has reached the back-side of the projectile nucleus,
a rarefaction wave starts to propagate inwards with the velocity of sound,
leading to expansion and cooling \cite{RiExpans}. This rarefaction wave can
be clearly seen in the lower left panel.

In the three-fluid model, the energy density profile looks qualitatively
different, cf.\ right column of fig.\ \ref{fig1}. One observes that the 
projectile nucleons cross the symmetry plane $z=0$, i.e.\ their
mean-free path in the target matter is non-vanishing (in one-fluid
hydrodynamics it follows from symmetry arguments that the baryon and energy
flow through the plane $z=0$ vanish). Consequently, no real discontinuity  
appears, the energy density profile is smooth (within the numerical
accuracy). The shock wave (if there is any) is smeared out considerably,
which is due to the finite mean-free path (which in turn is linked to   
the total $NN$ cross section). Also, the smeared-out shock wave obviously
leads to less compression and heating of the shocked matter as compared  
to the discontinuity in the one-fluid limit. The entropy production in the
two scenarios remains to be investigated.

\section{Kinetic equilibration of projectile and target nucleons}
\begin{figure}[htb]
%\vspace*{-1.0cm}   
\centerline{\hbox{ 
\psfig{figure=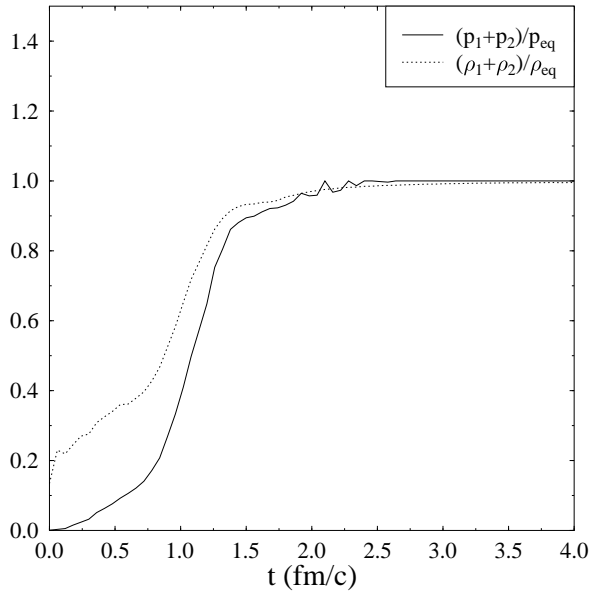,height=10cm,width=10cm}}}
\vspace*{-1cm}
\caption[]{
Ratio of the sum of projectile and target pressure (and baryon density)
to the equilibrium value as a function of CMS-time;
$Pb+Pb$-collisions ($b=0~fm$) at CERN-SPS.}
\label{fig1b}
\end{figure}
One important issue in ultrarelativistic heavy-ion collisions is
whether local thermal equilibrium is established via collisions
between the particles. To study this, we have calculated the ratio
of the sum of the individual pressures of the projectile and target
fluids, $p_1(t,\vec{x})+p_2(t,\vec{x})$, 
to the corresponding equilibrium pressure, $p_{eq}(t,\vec{x})$ (and similarly
for the baryon densities). This ratio was then averaged over the volume
where projectile and target overlap. This ratios approach unity if
the relative velocity (at the given space-time point) between the projectile
and target fluids becomes comparable to the thermal (resp.\ Fermi-)
velocities of the particles within each fluid (the exact definition is
given in ref.\ \cite{3fluid}). Thus, in one-fluid hydrodynamics it is
by definition always equal to one. In contrast, in the three-fluid model
(cf.\ fig.\ \ref{fig1b}) $\langle (p_1+p_2)/p_{eq}\rangle$ is close to zero
in the beginning and reaches unity only after $t_{CM}^{eq}\simeq2R_{Pb}/
\gamma_{CM}=1.5~fm/c$.
From this point on local kinetic equilibrium is established and the
one-fluid limit is valid for the subsequent expansion. Although this
time may look very short, it is significant for observables which are
sensitive to the very early reaction dynamics, e.g.\ directed flow and
``hard'' thermal photons (or dileptons). Due to the kinetic non-equilibrium
in the early stage, the conversion of initial longitudinal momentum
into transverse momentum is weaker than in the one-fluid limit.

\section{Directed Flow at SPS}
\begin{figure}[htb]
%\vspace*{-1.0cm}   
\centerline{\hbox{ 
\psfig{figure=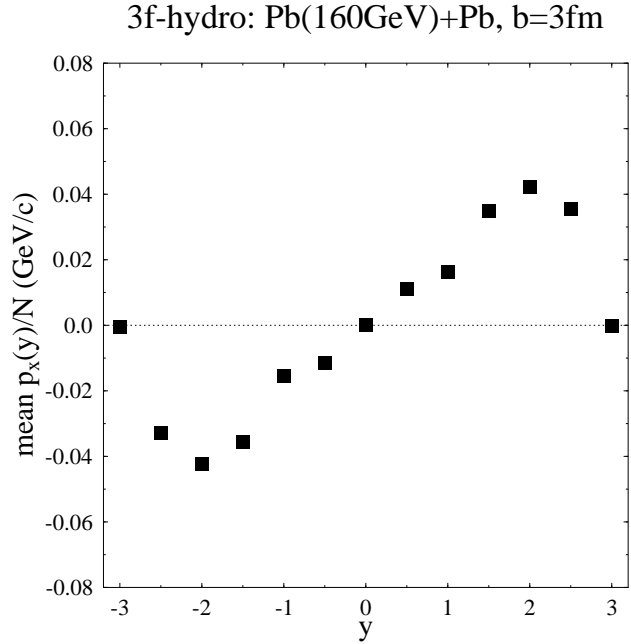,height=10cm,width=10cm}}} 
\vspace*{-1cm}
\caption[]{
Directed nucleon flow $\langle p_x^{dir}/N\rangle(y)$ in
$Pb+Pb$-collisions ($b=3~fm$) at CERN-SPS.}
\label{fig2}
\end{figure}
We already discussed in ref.\ \cite{3fluid} that in this model
kinetic equilibrium between projectile and target nucleons in the
central region occurs on the same time scale as is relevant for the onset
of directed nucleon flow. As a consequence, in the three-fluid model the 
directed nucleon flow is considerably lower and in particular less sensitive
to the equation-of-state of the nucleon fluids than in the one-fluid limit. 
In fig.\ \ref{fig2} we show our result for 
$\langle p_x^{dir}/N\rangle(y)$
for
$Pb+Pb$-collisions at CERN-SPS (in calculating this quantity we have
neglected thermal smearing of the nucleon momenta; for a definition see
ref.\ \cite{3fluid}). The maximum is on the order of $40-50~MeV$
and considerably lower than at AGS (due to the fact that the time 
scale is shorter by roughly a factor of three). Quantitative
comparisons to experimental data (directed flow was recently discovered in
such reactions \cite{Gutbrod}) may reveal whether our present model produces
enough directed flow or if the contribution of the pressure of the
produced particles (as mentioned above, rescattering of produced  
particles with the projectile and target nucleons is presently neglected
in our model) is essential. The produced particles are closer in
rapidity to the projectile nucleons than the target nucleons (and vice versa),
and thus might thermalize faster and exhibit significant pressure on them.
Directed nucleon flow at SPS could thus be more sensitive to the
equation-of-state of the produced particles than to that of the nucleons
themselves. This will be investigated in future work.

\section{Thermal photon emission from the third fluid}
\begin{figure}[htb]
\vspace*{-1.0cm}   
\centerline{\hbox{ 
\psfig{figure=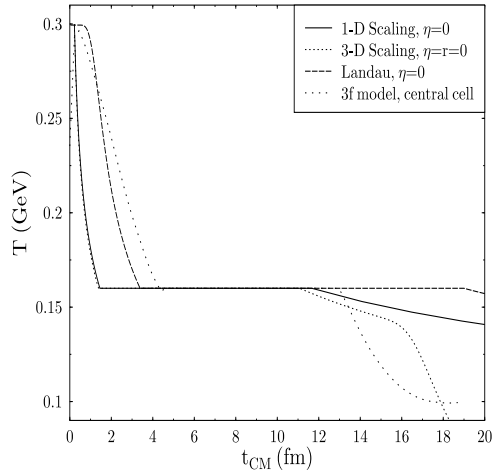,height=10cm,width=12cm}}} 
\vspace*{-1.9cm}
\caption[]{
Temperature in the central region as a function of time in various
hydrodynamical models, $Pb+Pb$-collisions ($b=0~fm$) at CERN-SPS.}
\label{fig3}
\end{figure}
In our model, in $Pb+Pb$-collisions at CERN-SPS the third fluid locally
reaches energy densities up to $\approx10~GeV/fm^3$ (for comparison: 
in the one-fluid limit and for an equation-of-state appropriate for an    
ultrarelativistic ideal gas, $p=\epsilon/3$, the central region
reaches an energy density of $\epsilon/\epsilon_0=4\gamma_{CM}^2-3$, where
$\epsilon_0\approx0.15~GeV/fm^3$ denotes the energy density of nuclear
matter in the ground state). We therefore employ an equation-of-state
for an ideal QGP (described within the MIT bagmodel) above $T_C=160~MeV$.
Below $T_C$, we assume that $\pi$, $\eta$, $\rho$ and $\omega$ are the  
most abundant particles. The two equations-of-state are matched by
Gibbs conditions for phase equilibrium, thus leading to a first-order phase
transition.

The temperature of the third fluid in the central cell is depicted in
fig.\ \ref{fig3}. We also compare to two other hydrodynamical models 
for the expansion of the midrapidity fluid. In scaling hydrodynamics 
one assumes that the longitudinal flow velocity is $v_z=z/t$
\cite{scaling} (independent of time), whereas the initial condition for the 
(one-dimensional) Landau expansion is $v=0$. In these latter models we assumed
an initial temperature of $300~MeV$ (for the boostinvariant expansion an
initial time of $\tau_0=0.22~fm$ was employed). The maximum temperature in the
three-fluid model is determined by the coupling terms between fluids
one and two (and, of course, by the equations-of-state). The produced
particles cool fastest in scaling hydrodynamics, while
in the case of a Landau expansion the temperature
is constant until the rarefaction waves reach the center.
In this latter case one finds the longest-lived
mixed phase, due to the fact that we assumed purely one-dimensional   
(longitudinal) expansion.

\begin{figure}[htb]
\vspace*{-1.0cm}   
\centerline{\hbox{ 
\psfig{figure=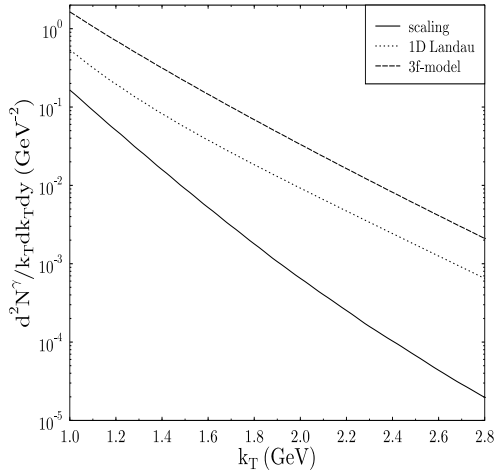,height=10cm,width=12cm}}}
\vspace*{-1.9cm}
\caption[]{
Transverse momentum distribution of direct photons at midrapidity
($y_{CM}^\gamma=0$) within
various hydrodynamical models, $Pb+Pb$-collisions ($b=0~fm$) at CERN-SPS.}
\label{fig4}
\end{figure}
The various cooling laws reflect in the transverse momentum  
distribution of direct photons \cite{Dum95a}, which can be produced in the
QGP- (by annihilation of quarks and antiquarks and Compton-scattering) and
hadronic phase (mainly by $\pi-\rho$ scattering) \cite{photons}.
The faster the cooling of the photon source, the steeper the slope of
the photon spectrum. Fitting the spectrum in the region
$2~GeV\le k_T\le3~GeV$ by an exponential distribution we find
$``T``=260~MeV$ in the three-fluid model and for the Landau expansion,
and $``T``=210~MeV$ for the scaling expansion.

\begin{figure}[htb]
\vspace*{-1.0cm}   
\centerline{\hbox{ 
\psfig{figure=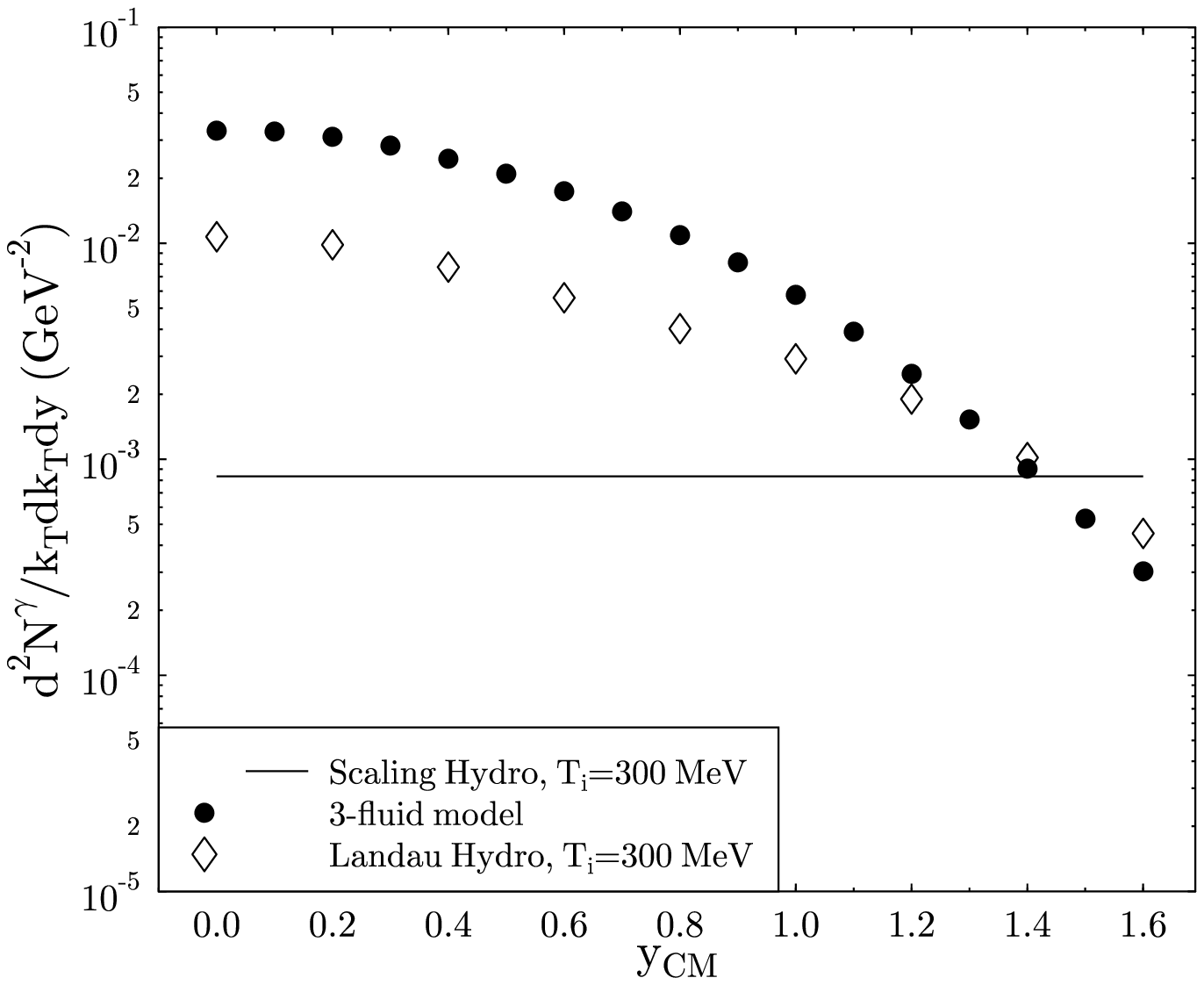,height=10cm,width=12cm}}}
\vspace*{-1.9cm}
\caption[]{
Rapidity distribution of direct photons with $k_T=2~GeV$ calculated within
various hydrodynamical models, $Pb+Pb$-collisions ($b=0~fm$) at CERN-SPS.}
\label{fig5}
\end{figure}
The rapidity distribution of thermal photons with transverse momenta
much larger than the maximal temperature of the photon source (e.g.\
$k_T=2~GeV$ at SPS) directly reflects the rapidity distribution of
the photon source at early times \cite{Dum95b}, cf.\ fig.\ \ref{fig5}.
This is due to the fact that ``hard'' photons do not thermalize,
in contrast to e.g.\ pions. Therefore, in the three-fluid model and in
the Landau-expansion case the photon rapidity distribution is strongly
peaked around midrapidity and is not proportional to the (squared) rapidity
distribution of the pions at freeze-out, as in scaling hydrodynamics. 

\section{Conclusions}
We discussed the compression of projectile and target and showed that
the shock discontinuities occuring in the one-fluid limit are smeared
out considerably in the three-fluid model. The energy density of the
baryons is less than in the one-fluid limit. Most of the energy loss
due to stopping goes into production of new particles, leading to the
creation of a very hot (temperatures up to $300~MeV$ at CERN-SPS) third
fluid. We compared the cooling of this third fluid in the three-fluid model
and two other simple expansion models, longitudinal Landau expansion and
longitudinally boostinvariant (plus cylindrically symmetric radial)
expansion, and showed how the various cooling laws reflect in the
transverse momentum distribution of thermal photons. The rapidity
distribution of thermal photons with transverse momenta much larger than
the maximal temperature measures the rapidity spread of the photon
source at early times (before acceleration leads to broadening).
Finally, in the three-fluid model the time scale for kinetic
equilibration of the nucleons in the central region is found to be on the order
of the Lorentz-contracted nuclear diameter, $2R/\gamma_{CM}$. This
leads to a less efficient conversion of initial longitudinal momentum
into transverse momentum as compared to the one-fluid limit. As a consequence,
the directed flow, which is also created in the early stage of the reaction,
is less than in the one-fluid limit.

\section*{Acknowledgement}
We thank I.\ Mishustin, L.\ Satarov, and D.H.\ Rischke for helpfull
discussions and comments.

\vfill\eject

\begin{thebibliography}{99}  
\bibitem{refppdNdy} V.~Blobel et al.: Nucl. Phys. B69 (1974) 454
\bibitem{2fluid} I.N.~Mishustin, V.N.~Russkikh, L.M.~Satarov:
Sov. J. Nucl. Phys. 48 (1988) 454; Nucl. Phys. A494 (1989) 595;\\
L.M.~Satarov: Sov. J. Nucl. Phys. 52 (1990) 264;\\
I.N. Mishustin, L.M. Satarov,  V.N. Russkikh: in ``Relativistic Heavy Ion
Physics'' (eds. D. Strottman and L.P. Csernai), vol. 1,
World Scientific (Singapore), 1991, p.179 and
Sov. J. Nucl. Phys. 54 (1991) 459
\bibitem{3fluid} J. Brachmann, A. Dumitru,
J.A. Maruhn, H. St\"ocker, W. Greiner, D.H. Rischke:
preprint UFTP-435/1997 (nucl-th/9703032), Nucl. Phys. A in print;\\
J. Brachmann, A. Dumitru, M. Bleicher, J.A. Maruhn, H. St\"ocker, W. Greiner:
Proc. of the XXXV Int. Winter Meeting Nucl. Phys., Bormio (Italy), 1997
(ed.: I. Iori) (nucl-th 9703044);\\
A. Dumitru: PhD thesis, Univ.\ Frankfurt, 1997
(http://www.th.physik.uni-frankfurt.de/${\tilde{}}$dumitru/lay.ps)
\bibitem{previous}
A.A.~Amsden, A.S.~Goldhaber, F.H.~Harlow,
J.R.~Nix: Phys. Rev. C17 (1978) 2080;\\  
L.P. Csernai et al.: Phys. Rev. C26 (1982) 149;\\
R.B. Clare, D. Strottman: Phys. Rep. 141 (1986) 177;\\
H.W. Barz, B. K\"ampfer: Phys. Lett. B206 (1988) 399  
\bibitem{Scheid} W. Scheid, R. Ligensa, W. Greiner: Phys. Rev. Lett.
21 (1968) 1479;\\
W. Scheid, W. Greiner: Z. Phys. 226 (1969) 364;\\
W. Scheid, H. M\"uller, W. Greiner: Phys. Rev. Lett. 32 (1974) 741
\bibitem{RiShock} D.H. Rischke, Y. P\"urs\"un, J.A. Maruhn:
Nucl. Phys. A595 (1995) 383; Erratum-ibid. A596 (1996) 717
\bibitem{Taub} A.M. Taub: Phys. Rev. 74 (1948) 328
\bibitem{SHASTA} J.P. Boris, D.L. Book:
J. Comput. Phys. 11 (1973) 38;\\
D.L. Book, J.P. Boris, K. Hain: J. Comput. Phys. 18 (1975) 248
\bibitem{RiExpans} D.H. Rischke, S. Bernard, J.A. Maruhn:
Nucl. Phys. A595 (1995) 346
\bibitem{Gutbrod} H.H. Gutbrod: private communication
\bibitem{scaling} K. Kajantie, L. McLerran: Phys. Lett. B119 (1982) 203; 
Nucl. Phys. B214 (1983) 261;\\
J.D. Bjorken: Phys. Rev. D27 (1983) 140;\\
K. Kajantie, R. Raitio, P.V. Ruuskanen: Nucl. Phys. B222 (1983) 152;\\
H. von Gersdorff, M. Kataja, L. McLerran, P.V. Ruuskanen:
Phys. Rev. D34 (1986) 794
\bibitem{Dum95a} 
D.K. Srivastava, B. Sinha: Phys. Rev. Lett. 73 (1994) 2421;\\
N. Arbex, U. Ornik, M. Pl\"umer, A. Timmermann,
R.M. Weiner: Phys. Lett. B345 (1995) 307;\\
J.J. Neumann, D. Seibert, G. Fai: Phys. Rev. C51 (1995) 1460;\\
A.~Dumitru, U.~Katscher, J.A.~Maruhn, H.~St\"ocker,
W.~Greiner, D.H.~Rischke: Phys. Rev. C51 (1995) 2166;\\
J. Sollfrank,  P. Huovinen, M. Kataja, P.V. Ruuskanen,
M. Prakash, R. Venugopalan: Phys. Rev. C55 (1997) 392
\bibitem{photons} J. Kapusta, P. Lichard, D. Seibert: Phys. Rev.
D44 (1991) 2774
\bibitem{Dum95b} A.~Dumitru, U.~Katscher, J.A.~Maruhn, H.~St\"ocker,
W.~Greiner, D.H.~Rischke: Z. Phys. A353 (1995) 187
\end{thebibliography}
\end{document}